# DIAGNOSIS OF DIABETES USING CLASSIFICATION MINING TECHNIQUES

Aiswarya Iyer, S. Jeyalatha and Ronak Sumbaly

Department of Computer Science, BITS Pilani Dubai, United Arab Emirates

***ABSTRACT***

*Diabetes has affected over 246 million people worldwide with a majority of them being women. According to the WHO report, by 2025 this number is expected to rise to over 380 million. The disease has been named the fifth deadliest disease in the United States with no imminent cure in sight. With the rise of information technology and its continued advent into the medical and healthcare sector, the cases of diabetes as well as their symptoms are well documented. This paper aims at finding solutions to diagnose the disease by analyzing the patterns found in the data through classification analysis by employing Decision Tree and Naïve Bayes algorithms. The research hopes to propose a quicker and more efficient technique of diagnosing the disease, leading to timely treatment of the patients.*

***KEYWORDS***

*Classification, Data Mining, Decision Tree, Diabetes and Naïve Bayes.*

## 1. INTRODUCTION

Effects of diabetes have been reported to have a more fatal and worsening impact on women than on men because of their lower survival rate and poorer quality of life. WHO reports state that almost one – third of the women who suffer from diabetes have no knowledge about it. The effect of diabetes is unique in case of mothers because the disease is transmitted to their unborn children. Strokes, miscarriages, blindness, kidney failure and amputations are just some of the complications that arise from this disease. For the purposes of this paper, the analyses of diabetes cases have been restricted to pregnant women.

Generally a person is considered to be suffering from diabetes, when blood sugar levels are above normal (4.4 to 6.1 mmol/L) [1]. Pancreas present in the human body produces insulin, a hormone that is responsible to help glucose reach each cell of the body. A diabetic patient essentially has low production of insulin or their body is not able to use the insulin well. There are three main types of diabetes, viz. Type 1, Type 2 and Gestational [2]. Type 1 – The disease manifest as an autoimmune disease occurring at a very young age of below 20 years. In this type of diabetes, the pancreatic cells that produce insulin have been destroyed. Type 2 - Diabetes is in the state when the various organs of the body become insulin resistant, and this increases the demand for insulin. At this point, pancreas doesn't make the required amount of insulin. Gestational diabetes tends to occur in pregnant women, as the pancreas don't make sufficient amount of insulin. All these types of diabetes need treatment and if they are detected at an early state, one can avoid the complications associated with them.

Nowadays, large amount of information is collected in the form of patient records by the hospitals. Knowledge discovery for predictive purposes is done through data mining, which is a analysis technique that helps in proposing inferences. This method helps in decision-making through algorithms from large amounts of data generated by these medical centers. Considering the importance of early medical diagnosis of this disease, data mining techniques can be applied to help the women in detection of diabetes at an early stage and treatment, which may help in avoiding complications.

The paper focuses on diabetes recorded in pregnant women. In this paper, Decision Tree and Naïve Bayes algorithm have been employed on a pre-existential dataset to predict whether diabetes is recorded or not in a patient. Results from both the algorithms have been compared and presented. Several other models have been formulated over the years that are used for diabetes prediction. Some of this related work is discussed in Section 5 of this paper.

*Structure of the Paper*

The remainder of the paper is organized as follows: In Section 2, the objectives of this paper are presented. In Section 3, the various acronyms used in this paper have been given. An overview of Diabetes is presented in Section 4. Section 5 deals with summarizing the related work done under the scope of this paper. The fundamentals of data mining techniques used along with its standard tasks are presented in Section 6. A Decision Tree and Naïve Bayes model for diabetes prediction is presented in Section 7. Section 8 discusses the results and analysis of the model. The paper is concluded in Section 9.

## 2. OBJECTIVES

The present work is intended to meet the following objectives:

1. Present a Decision Tree and Naïve Bayes model for diabetes prediction in pregnant women.
2. Summarize Diabetes – types, risk factors, symptoms and diagnosis.
3. Identify and discuss the field's benefits to the society along with effective application.

## 3. ACRONYMS

ACO : Ant Colony Optimization
ANN : Artificial Neural Network
CSV : Comma Separated Values
FPG : Fasting Plasma Glucose
GA : Genetic Algorithm
IG : Information Gain
OGTT : Oral Glucose Tolerance Test

## 4. OVERVIEW OF DIABETES

### 4.1. Diabetes

Diabetes is a disease that occurs when the insulin production in the body is inadequate or the body is unable to use the produced insulin in a proper manner, as a result, this leads to high blood glucose. The body cells break down the food into glucose and this glucose needs to be transported to all the cells of the body. The insulin is the hormone that directs the glucose that is produced by breaking down the food into the body cells. Any change in the production of insulin leads to an increase in the blood sugar levels and this can lead to damage to the tissues and failure of the organs. Generally a person is considered to be suffering from diabetes, when blood sugar levels are above normal (4.4 to 6.1 mmol/L). There are three main types of diabetes, viz. Type 1, Type 2 and Gestational.

## 4.2. Types of Diabetes

The three main types of diabetes are described below:

1. **Type 1** – Though there are only about 10% of diabetes patients have this form of diabetes, recently, there has been a rise in the number of cases of this type in the United States. The disease manifest as an autoimmune disease occurring at a very young age of below 20 years hence also called juvenile-onset diabetes. In this type of diabetes, the pancreatic cells that produce insulin have been destroyed by the defence system of the body. Injections of insulin along with frequent blood tests and dietary restrictions have to be followed by patients suffering from Type 1 diabetes.

2. **Type 2** – This type accounts for almost 90% of the diabetes cases and commonly called the adult-onset diabetes or the non-insulin dependent diabetes. In this case the various organs of the body become insulin resistant, and this increases the demand for insulin. At this point, pancreas doesn't make the required amount of insulin. To keep this type of diabetes at bay, the patients have to follow a strict diet, exercise routine and keep track of the blood glucose. Obesity, being overweight, being physically inactive can lead to type 2 diabetes. Also with ageing, the risk of developing diabetes is considered to be more. Majority of the Type 2 diabetes patients have border line diabetes or the Pre-Diabetes, a condition where the blood glucose levels are higher than normal but not as high as a diabetic patient.

3. **Gestational diabetes** – is a type of diabetes that tends to occur in pregnant women due to the high sugar levels as the pancreas don't produce sufficient amount of insulin. Taking no treatment can lead to complications during childbirth. Controlling the diet and taking insulin can control this form of diabetes.

All these types of diabetes are serious and need treatment and if they are detected at an early state, one can avoid the complications associated with them.

## 4.3. Symptoms, Diagnosis and Treatment

The common symptoms of a person suffering from diabetes are:

- Polyuria (frequent urination)
- Polyphagia (excessive hunger)
- Polydipsia (excessive thirst)
- Weight gain or strange weight loss
- Healing of wounds is not quick, blurred vision, fatigue, itchy skin, etc.

Urine test and blood tests are conducted to detect diabetes by checking for excess body glucose. The commonly conducted tests for determining whether a person has diabetes or not are

- A1C Test
- Fasting Plasma Glucose (FPG) Test
- Oral Glucose Tolerance Test (OGTT).

Though both Type 1 and Type 2 diabetes cannot be cured they can be controlled and treated by special diets, regular exercise and insulin injections. The complications of the disease include neuropathy, foot amputations, glaucoma, cataracts, increased risk of kidney diseases and heart attack and stroke and many more.

The earlier diagnosis of diabetes, risk of the complications can be dodged. Hence a faster method of predicting the disease has been presented in this paper.

## 5. RELATED WORK

Design of prediction models for diabetes diagnosis has been an active research area for the past decade. Most of the models found in literature are based on clustering algorithms and artificial neural networks (ANNs).

In [5] the authors have employed three techniques namely: EM algorithm, H-means+ clustering and Genetic Algorithm (GA), for the classification of the diabetic patients. The performance for H-means+ proved to be better than others when all the similar symptoms were grouped into clusters using these algorithms. A study conducted in [6] intended to discover the hidden knowledge from a particular dataset to improve the quality of health care for diabetic patients. In [7] Fuzzy Ant Colony Optimization (ACO) was used on the Pima Indian Diabetes dataset to find set of rules for the diabetes diagnosis.

The paper [8] approached the aim of diagnoses by using ANNs and demonstrated the need for preprocessing and replacing missing values in the dataset being considered. Through the modified training set, a better accuracy was achieved with lesser time required for training the set. Finally in [9] a neural network model for prediction of diabetes based on 13 early symptoms of the disease was created with implementation using MATLAB.

However, nobody established a classification model based on probability and feature selection. In all the related work analytical techniques have been employed to produce reliable results but generally the methods are time consuming since most employed a weighted approach.

Hence, there is a requirement of a model that can be developed easily providing reliable, faster and cost effective methods to provide information of the probability of a patient to have diabetes. In the present work, an attempt is made to analyze the diabetes parameters and to establish a probabilistic relation between them using Naïve Bayes and Decision Tree approach. For the purpose of analysis the models are tested depending on the percentage of correctly classified instances in the dataset.

## 6. OVERVIEW OF METHODOLOGIES

The present work intends to create a mining model based on two classification algorithms in order to provide a simpler solution to the problem of diagnosis of diabetes in women. The results have been analyzed using statistical methods and are presented in the Section 6.1 and Section 6.2.

### 6.1. Decision Trees

Decision tree [3] is a tree structure, which is in the form of a flowchart. It is used as a method for classification and prediction with representation using nodes and internodes. The root and internal nodes are the test cases that are used to separate the instances with different features. Internal nodes themselves are the result of attribute test cases. Leaf nodes denote the class variable. Figure 1. shows a sample decision tree structure.

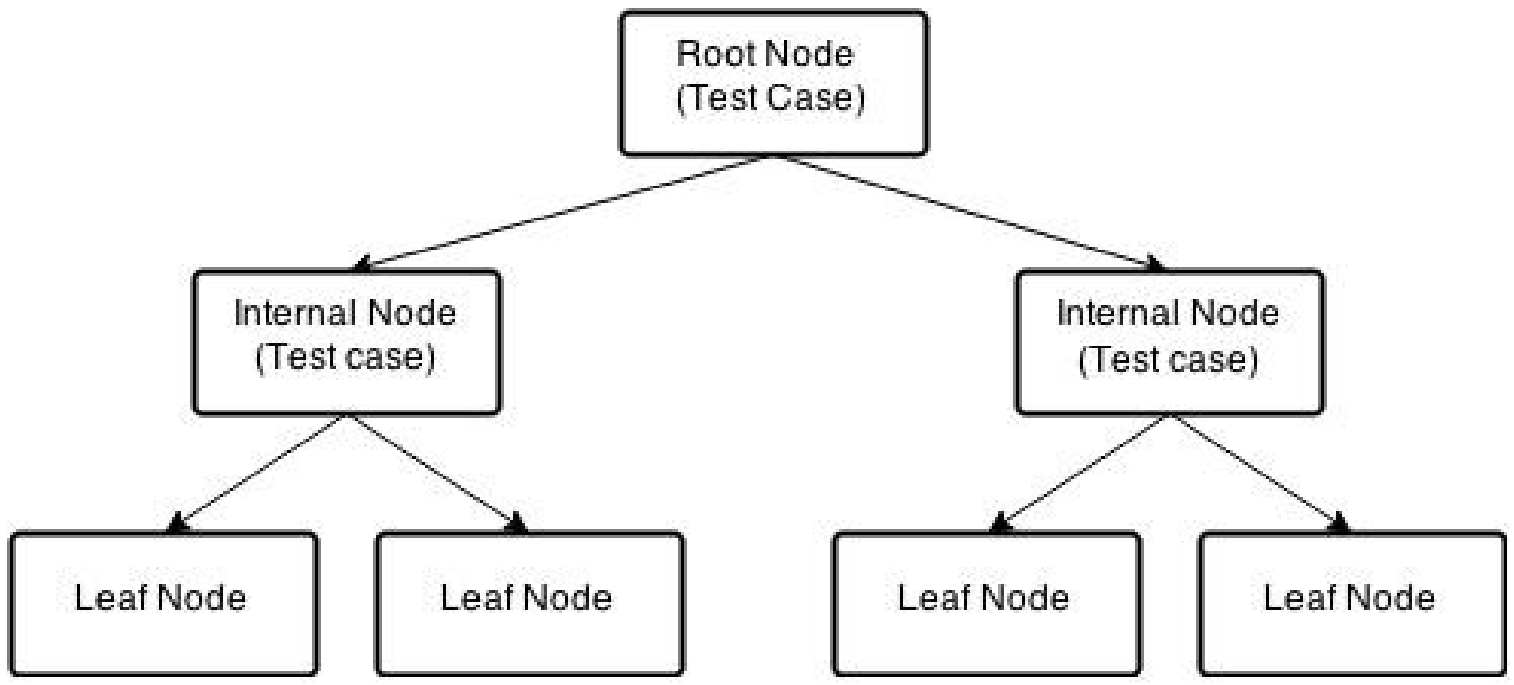

Figure 1. Sample Decision Tree Structure

Decision tree provides a powerful technique for classification and prediction in Diabetes diagnosis problem. Various decision tree algorithms are available to classify the data, including ID3, C4.5, C5, J48, CART and CHAID. In this paper, J48 decision tree algorithm [10] has been chosen to establish the model. Each node for the decision tree is found by calculating the highest information gain for all attributes and if a specific attribute gives an unambiguous end product (explicit classification of class attribute), the branch of this attribute is terminated and target value is assigned to it.

### 6.2. Naïve Bayes

The Naïve Bayes Algorithm is a probabilistic algorithm that is sequential in nature, following steps of execution, classification, estimation and prediction. For finding relations between the diseases, symptoms and medications, there are various data mining existing solution, but these algorithms have their own limitations; numerous iterations, binning of the continuous arguments, high computational time, etc. Naïve Bayes overcomes various limitations including omission of complex iterative estimations of the parameter and can be applied on a large dataset in real time. The algorithm works on the simple Naïve Bayes formula shown in Fig 2.

$$\underset{P(c|x)}{\text{Posterior Probability}} = \frac{\underset{P(x|c)}{\text{Likelihood}} \times \underset{P(c)}{\text{Class Prior Probability}}}{\underset{P(x)}{\text{Predictor Prior Probability}}}$$

Figure 2. Naïve Bayes Formula

## 7. METHODOLOGIES

### 7.1. Dataset Description and Pre-Processing

The paper explores the aspect of Decision Tree and Naïve Bayes Classifier as Data Mining techniques in determining diabetes in women. The main objective is to forecast if the patient has been affected by diabetes using the data mining tools by using the medical data available. The classification type of data mining has been applied to the Pima Indians Diabetes Database of National Institute of Diabetes and Digestive and Kidney Diseases. Table 1 shows a brief description of the dataset that is being considered.

Table 1. Dataset Description.

| Dataset | No. of Attributes | No. of Instances |
|---|---|---|
| Pima Indians Diabetes Database of National Institute of Diabetes and Digestive and Kidney Diseases | 8 | 768 |

The attributes descriptions are shown in Table 2 below.

Table 2. Attribute Description.

| Attribute | Relabeled values |
|---|---|
| 1. Number of times pregnant | Preg |
| 2. Plasma glucose concentration | Plas |
| 3. Diastolic blood pressure (mm Hg) | Pres |
| 4. Triceps skin fold thickness (mm) | Skin |
| 5. 2-Hour serum insulin | Insu |
| 6. Body mass index (kg/m$^2$) | Mass |
| 7. Diabetes pedigree function | Pedi |
| 8. Age (years) | Age |
| 9. Class Variable (0 or 1) | Class |

Pre-processing and transformation of the dataset are done using WEKA tools [11].

Transformation steps include:

- Replacing missing values, and
- Normalization of values.

The latter makes it easy to use the dataset, as the range of the variables are restricted from 0 to 1. Feature selection has been employed using the CfsSubsetEval algorithm, and the attributes obtained after execution are as follows:

1. Plasma glucose concentration
2. Body mass index (kg/m$^2$)
3. Diabetes pedigree function
4. Age (years)
5. Class Variable (nominal) - Determines if the person has diabetes or not

The descriptive statistics of the dataset are presented in Table 3. Since the parameters are normalized the range of all are in the range 0 to 1.

Table 3. Descriptive Statistics of Transformed Dataset

| Parameter | Minimum | Maximum | Mean | Std. Deviation |
|---|---|---|---|---|
| Plas | 0 | 1 | 0.608 | 0.161 |
| Mass | 0 | 1 | 0.477 | 0.117 |
| Pedi | 0 | 1 | 0.168 | 0.141 |
| Age | 0 | 1 | 0.204 | 0.196 |

Figure. 3 shows the distribution of the class attribute in the dataset.

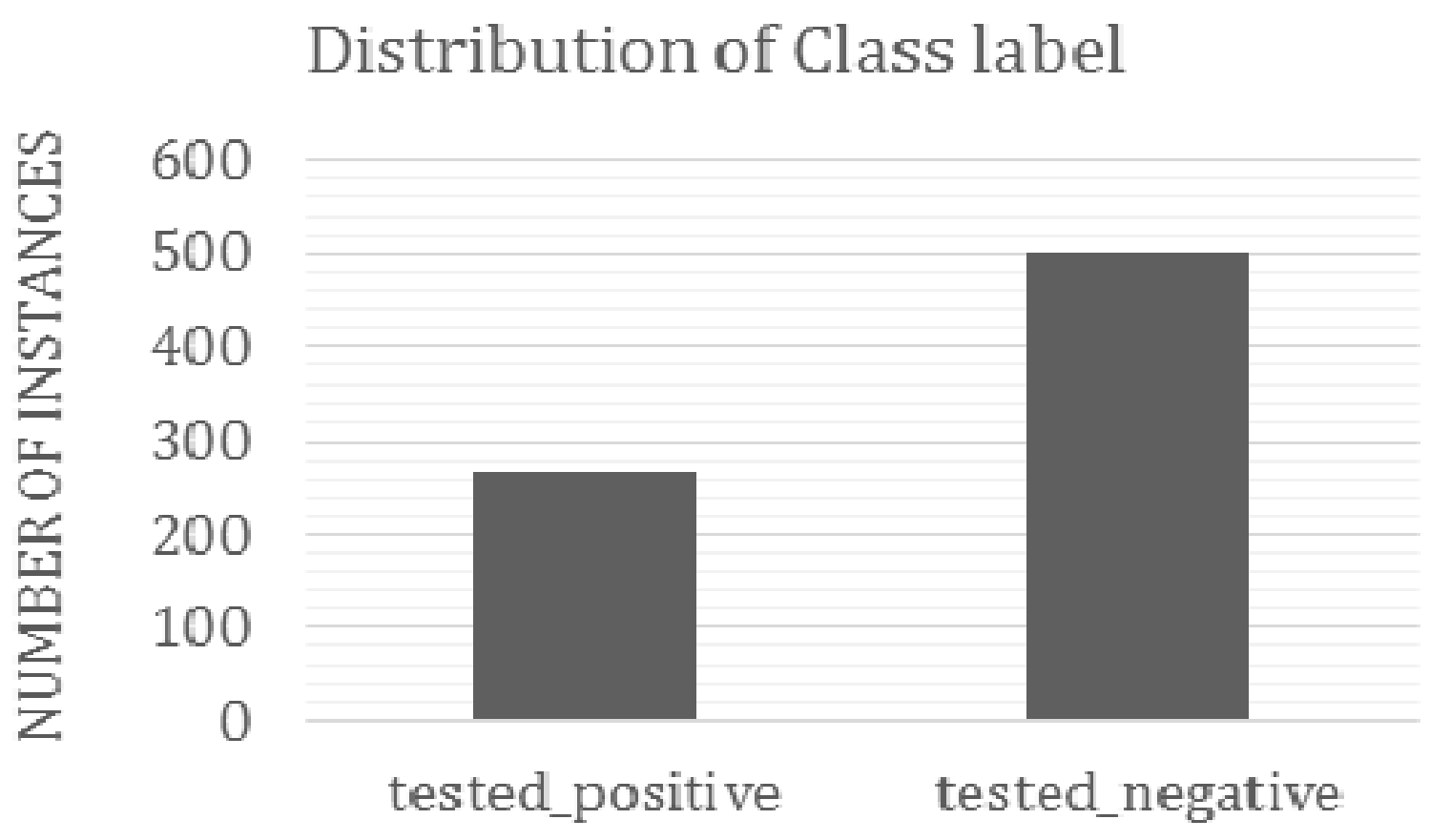

Figure 3. Class Attribute Distribution

## 7.2. Proposed Data Model

In this paper two algorithms namely, J48 (decision tree algorithm) and Naïve Bayes, have been used to create the model for diagnosis. The data was divided into training set and test set by the cross-validation technique and percentage split technique.

10-fold cross validation is used to prepare training and test data. After data pre-processing (CSV format), the J48 algorithm is employed on the dataset using WEKA (Java Toolkit for various data mining technique) after which data are divided into “tested-positive” or “tested-negative” depending on the final result of the decision tree that is constructed. The algorithm for conducting the procedure is as follows:

---

**ALGORITHM:** DIABETES_ALGO

---

***INPUT*:** Pima Indians Diabetes Database of National Institute of Diabetes and Digestive and Kidney Diseases dataset pre-processed in CSV format.

***OUTPUT*:** J48 Decision Tree Predictive Model with leaf node either tested-positive or tested-negative and Naïve Bayes Prediction Results.

**PROCEDURE:**

1. The dataset is pre-processed using WEKA tools. Following operations are performed on the dataset
    a. Replace Missing Values and
    b. Normalization of values.
2. Processed dataset is passed through feature selection wherein sets of attributes are deleted from the dataset.
3. The final processed dataset is uploaded in WEKA

4. The J48 Decision Tree and Naïve Bayes algorithm are employed.

5. For purposes of the algorithms, Cross-Validation and Percentage Split techniques are applied for model creation.

Both models analyzed on the basis of correctly classified instances.

Figure 4. shows the flow of the research conducted to construct the model

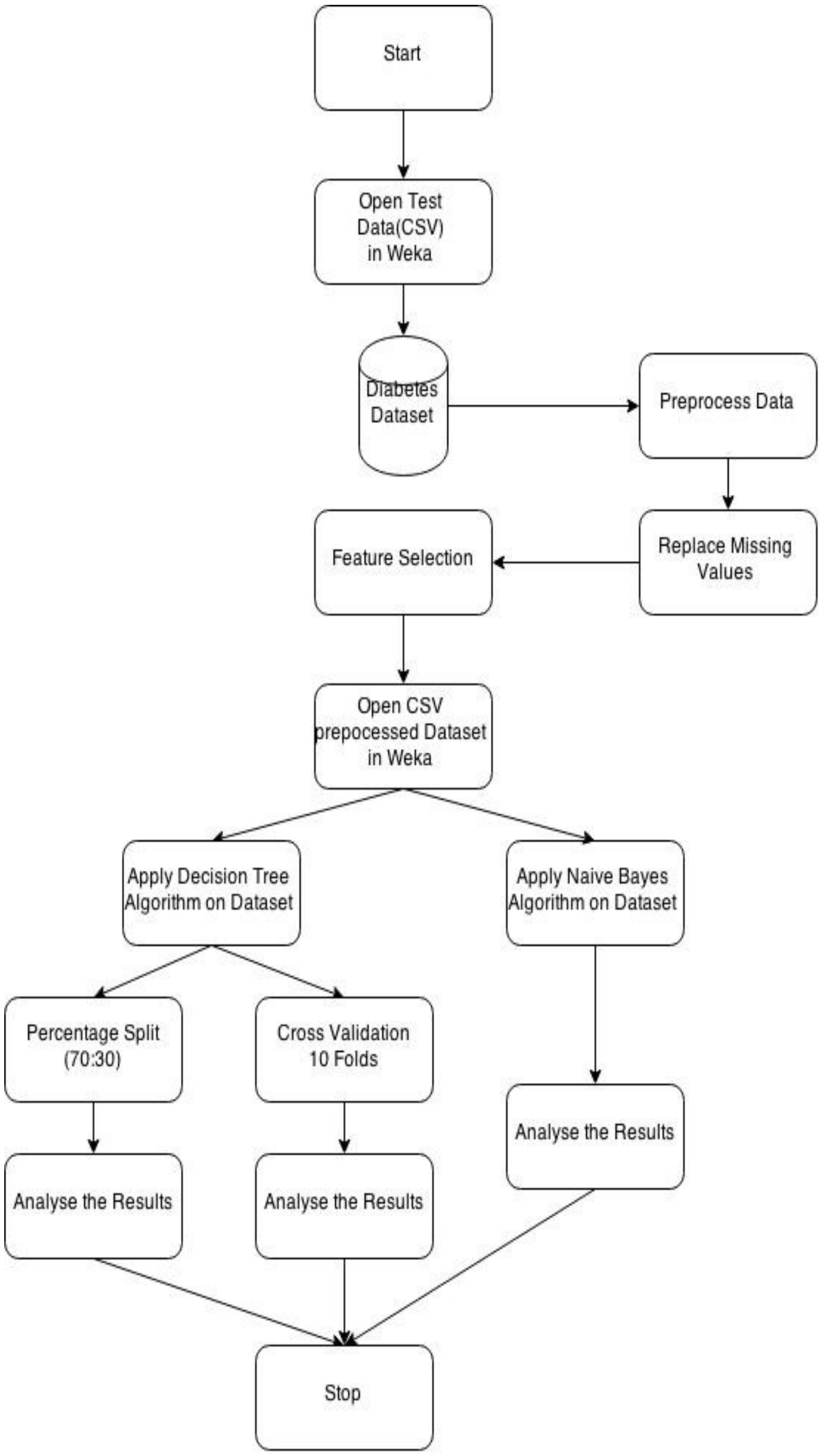

Figure 4. Flowchart depicting Model Creation

## 8. RESULTS AND ANALYSIS

### 8.1. J48 Decision Tree

Decision tree J48 implements Quinlan's C4.5 algorithm [10] for generating pruned tree. The tree generated by J48 can be used for classification of whether a patient has tested positive or negative for diabetes. The data mining technique uses the concept of information gain. Each attribute of the data is used to make a decision by splitting the data into smaller modules.

It examines normalized information gain (IG) [12] (difference in entropy) that results from choosing an attribute as a split point. The highest normalized IG is used at the root of the tree. The procedure is repeated until the leaf node is created for the tree specifying the class attribute that is chosen. Figure. 5 shows the J48 pruned tree that was generated by WEKA.

```
J48 pruned tree
------------------

plas <= 0.638191
|   mass <= 0.393443: tested_negative (132.0/3.0)
|   mass > 0.393443
|   |   age <= 0.116667: tested_negative (180.0/22.0)
|   |   age > 0.116667
|   |   |   plas <= 0.497487: tested_negative (55.0/10.0)
|   |   |   plas > 0.497487
|   |   |   |   pedi <= 0.206234: tested_negative (84.0/34.0)
|   |   |   |   pedi > 0.206234: tested_positive (34.0/9.0)
plas > 0.638191
|   mass <= 0.445604
|   |   plas <= 0.728643: tested_negative (41.0/6.0)
|   |   plas > 0.728643
|   |   |   age <= 0.066667: tested_negative (4.0)
|   |   |   age > 0.066667
|   |   |   |   age <= 0.666667
|   |   |   |   |   mass <= 0.403875: tested_positive (12.0/1.0)
|   |   |   |   |   mass > 0.403875
|   |   |   |   |   |   pedi <= 0.135781: tested_positive (11.0/4.0)
|   |   |   |   |   |   pedi > 0.135781: tested_negative (4.0)
|   |   |   |   age > 0.666667: tested_negative (4.0)
|   mass > 0.445604
|   |   plas <= 0.788945
|   |   |   pedi <= 0.149018
|   |   |   |   mass <= 0.678092: tested_negative (50.0/21.0)
|   |   |   |   mass > 0.678092: tested_positive (7.0)
|   |   |   pedi > 0.149018: tested_positive (58.0/16.0)
|   |   plas > 0.788945: tested_positive (92.0/12.0)

Number of Leaves  :     15

Size of the tree :      29
```

Figure 5. J48 Pruned Tree

***Classifier Output***

***Based on Cross-Validation Technique***

The J48 algorithm gives the following correctness results for the given dataset.

Table 4. Performance Results from J48 Classification Algorithm – Cross Validation

| | **No. of Instances** | **Percentage** |
|---|---|---|
| Correctly Classified Instances | 575 | 74.8698 % |
| Incorrectly Classified Instances | 193 | 25.1302 % |

The other results are presented in Table 5.

Table 5. Other Results of J48 Classification Algorithm – Cross Validation

| | |
|---|---|
| **Kappa statistic** | 0.4245 |
| **Mean absolute error** | 0.4245 |
| **Root mean squared error** | 0.4216 |
| **Relative absolute error** | 69.4357 % |
| **Root relative squared error** | 88.4504 % |
| **Total Number of Instances** | 768 |

*Terminologies of Test Statistics:*

1. Kappa Statistic: is a metric that compares Observed Accuracy with Expected Accuracy (random chance).
2. Mean Absolute Error: average of the absolute error between observed and forecasted value.
3. Root Mean Squared Error: measure of the differences between value (Sample and population values) predicted by a model or an estimator and the values actually observed
4. Relative Absolute Error: ratio of the absolute error of the measurement to the accepted measurement

Table 6 shows the confusion matrix of the J48 Decision Tree.

Table 6. Confusion Matrix – Cross Validation

| | *A - tested_positive* | *B - tested_negative* |
|---|---|---|
| ***A - tested_positive*** | *149 (i)* | *119 (ii)* |
| ***B - tested_negative*** | *74 (iii)* | *426 (iv)* |

In the table, the values represent following:

i. : Number of correct forecasts that the instance tested positive
ii. : Number of incorrect forecasts that the instance tested negative
iii. : Number of incorrect forecasts that the instance tested positive
iv. : Number of correct forecasts that the instance tested negative

***Based on Percentage Split (70:30) Technique***

Since a 70:30 percentage split was applied on the dataset 230 of the instances were used as the test dataset while the rest of were using for training the model. The J48 algorithm gives the following correctness results for the given dataset.

Table 7. Performance Results from J48 Classification Algorithm – Percentage Split

| | **No. of Instances** | **Percentage** |
|---|---|---|
| Correctly Classified Instances | 177 | 76.9565 % |
| Incorrectly Classified Instances | 53 | 23.0435 % |

The other results are presented in Table 8.

Table 8. Other Results of J48 Classification Algorithm – Percentage Split

| | |
|---|---|
| **Kappa statistic** | 0.4742 |
| **Mean absolute error** | 0.3374 |
| **Root mean squared error** | 0.4029 |
| **Relative absolute error** | 75.0763 % |
| **Root relative squared error** | 86.3392 % |
| **Total Number of Instances** | 230 |

Confusion matrix for the J48 algorithm is shown in Table 9.

Table 9. Confusion Matrix – Percentage Split

| | *A - tested_positive* | *B - tested_negative* |
|---|---|---|
| ***A - tested_positive*** | *48* | *24* |
| ***B - tested_negative*** | *29* | *129* |

## 8.2. Naïve Bayes

Only percentage split technique was applied for the Naïve Bayes algorithm. The results are shown in Table 10.

***Classifier Output***

***Based on Percentage Split (70:30) Technique***

The Naïve Bayes Algorithm gives the following correctness results for the given dataset:

Table 10 .Performance Results from Naïve Bayes – Percentage Split

| | **No. of Instances** | **Percentage** |
|---|---|---|
| Correctly Classified Instances | 183 | 79.5652 % |
| Incorrectly Classified Instances | 47 | 20.4348 % |

The other results are presented in Table 11.

Table 11. Other Results of Naïve Bayes Classification Algorithm – Percentage Split

| | |
|---|---|
| **Kappa statistic** | 0.5081 |
| **Mean absolute error** | 0.2884 |
| **Root mean squared error** | 0.381 |
| **Relative absolute error** | 64.175 % |
| **Root relative squared error** | 81.6565 % |
| **Total Number of Instances** | 230 |

Confusion matrix for the Naïve Bayes Algorithm is shown in Table 12.

Table 12. Confusion Matrix – Naïve Bayes

| | *A - tested_positive* | *B - tested_negative* |
|---|---|---|
| ***A - tested_positive*** | *44* | *28* |
| ***B - tested_negative*** | *19* | *139* |

## 8.3. Analysis

From the results obtained, both the methods have a comparatively small difference in error rate, though the percentage split of 70:30 for Naïve Bayes technique gives the least error rate as compared to other two J48 implementations. Both the models are efficient in the diagnosis of diabetes using the percentage split of 70:30 of the data set. A developed model for diagnosis of diabetes will require more training data for creation and testing.

## 9. CONCLUSIONS

The automatic diagnosis of diabetes is an important real-world medical problem. Detection of diabetes in its early stages is the key for treatment. This paper shows how Decision Trees and Naïve Bayes are used to model actual diagnosis of diabetes for local and systematic treatment, along with presenting related work in the field. Experimental results show the effectiveness of the proposed model. The performance of the techniques was investigated for the diabetes diagnosis problem. Experimental results demonstrate the adequacy of the proposed model.

In future it is planned to gather the information from different locales over the world and make a more precise and general prescient model for diabetes conclusion. Future study will likewise focus on gathering information from a later time period and discover new potential prognostic elements to be incorporated. The work can be extended and improved for the automation of diabetes analysis.

## REFERENCES

[1] National Diabetes Information Clearinghouse (NDIC), http://diabetes.niddk.nih.gov/dm/pubs/type1and2/#signs

[2] Global Diabetes Community, http://www.diabetes.co.uk/diabetes_care/blood-sugar-level-ranges.html

[3] Jiawei Han and Micheline Kamber, "Data Mining Concepts and Techniques", Morgan Kauffman Publishers, 2001

[4] S. Kumari and A. Singh, "A Data Mining Approach for the Diagnosis of Diabetes Mellitus", Proceedings of Seventh lnternational Conference on Intelligent Systems and Control, 2013, pp. 373-375

[5] C. M. Velu and K. R. Kashwan, "Visual Data Mining Techniques for Classification of Diabetic Patients", 3rd IEEE International Advance Computing Conference (IACC), 2013

[6] Sankaranarayanan.S and Dr Pramananda Perumal.T, "Predictive Approach for Diabetes Mellitus Disease through Data Mining Technologies", World Congress on Computing and Communication Technologies, 2014, pp. 231-233

[7] Mostafa Fathi Ganji and Mohammad Saniee Abadeh, "Using fuzzy Ant Colony Optimization for Diagnosis of Diabetes Disease", Proceedings of ICEE 2010, May 11-13, 2010

[8] T.Jayalakshmi and Dr.A.Santhakumaran, "A Novel Classification Method for Diagnosis of Diabetes Mellitus Using Artificial Neural Networks", International Conference on Data Storage and Data Engineering, 2010, pp. 159-163

[9] Sonu Kumari and Archana Singh, "A Data Mining Approach for the Diagnosis of Diabetes Mellitus", Proceedings of71hlnternational Conference on Intelligent Systems and Control (ISCO 2013)

[10] Neeraj Bhargava, Girja Sharma, Ritu Bhargava and Manish Mathuria, Decision Tree Analysis on J48 Algorithm for Data Mining. Proceedings of International Journal of Advanced Research in Computer Science and Software Engineering, Volume 3, Issue 6, June 2013.

[11] Michael Feld, Dr. Michael Kipp, Dr. Alassane Ndiaye and Dr. Dominik Heckmann "Weka: Practical machine learning tools and techniques with Java implementations"

[12] White, A.P., Liu, W.Z.: Technical note: Bias in information-based measures in decision tree induction. Machine Learning 15(3), 321–329 (1994)

## AUTHORS

**Ms. Aiswarya Iyer** is currently pursuing B.E (Hons) from Birla Institute of Technology & Science (BITS), Pilani in the department of Computer Science in Dubai, UAE

**Dr. S Jeyalatha** is currently a Senior Lecturer at Birla Institute of Technology & Science (BITS), Pilani in the department of Computer Science in Dubai, UAE

**Mr. Ronak Sumbaly** is currently pursuing B.E (Hons) from Birla Institute of Technology & Science (BITS), Pilani in the department of Computer Science in Dubai, UAE